\documentclass[12pt,prl,showpacs,hyperref,superscriptaddress]{revtex4}
\pdfoutput=1
\usepackage{color}
\usepackage{amsfonts}
\usepackage{amsmath}
\usepackage{amssymb}
\usepackage{graphicx}
\usepackage{subfigure}
\usepackage{psfrag}
\providecommand{\U}[1]{\protect\rule{.1in}{.1in}}
\providecommand{\U}[1]{\protect\rule{.1in}{.1in}}

\begin{document}

\def\blue#1{{\color{blue}{#1}}}

\title{Two-dimensional Bose and Fermi gases beyond weak coupling}

\author{Guilherme Fran\c ca}
\affiliation{
Department of Physics, Cornell University, Ithaca, NY}

\author{Andr\'e LeClair}
\affiliation{
Department of Physics, Cornell University, Ithaca, NY}

\author{Joshua Squires}
\affiliation{
Department of Physics, Cornell University, Ithaca, NY}

\begin{abstract}
\qquad

Using a formalism based on the two-body S-matrix we study two-dimensional Bose and Fermi gases with both attractive and repulsive interactions. Approximate analytic expressions, valid at weak coupling and beyond, are developed and applied to the Berezinskii-Kosterlitz-Thouless (BKT) transition. We successfully recover the correct logarithmic functional form of the critical chemical potential and density for the Bose gas. For fermions, the BKT critical temperature is calculated in BCS and BEC regimes through consideration of Tan's contact.

\end{abstract}

\pacs{05.45.Mt, 03.75.Mn, 03.67.-a}
\maketitle

%
%
%
\def\oti{{\otimes}}
\def\lb{ \left[ }
\def\rb{ \right]  }
\def\tilde{\widetilde}
\def\bar{\overline}
\def\hat{\widehat}
\def\*{\star}
\def\[{\left[}
\def\]{\right]}
\def\({\left(}
\def\BL{\Bigr(}
\def\){\right)}
\def\BR{\Bigr)}
\def\BBL{\lb}
\def\BBR{\rb}
\def\zb{{\bar{z} }}
\def\zbar{{\bar{z} }}
\def\frac#1#2{{#1 \over #2}}
\def\inv#1{{1 \over #1}}
\def\half{{1 \over 2}}
\def\d{\partial}
\def\der#1{{\partial \over \partial #1}}
\def\dd#1#2{{\partial #1 \over \partial #2}}
\def\vev#1{\langle #1 \rangle}
\def\bra#1{{\langle #1 |  }}
\def\ket#1{ | #1 \rangle}
\def\rvac{\hbox{$\vert 0\rangle$}}
\def\lvac{\hbox{$\langle 0 \vert $}}
\def\2pi{\hbox{$2\pi i$}}
\def\e#1{{\rm e}^{^{\textstyle #1}}}
\def\grad#1{\,\nabla\!_{{#1}}\,}
\def\dsl{\raise.15ex\hbox{/}\kern-.57em\partial}
\def\Dsl{\,\raise.15ex\hbox{/}\mkern-.13.5mu D}

\def\th{\theta}		
\def\Th{\Theta}
\def\ga{\gamma}		
\def\Ga{\Gamma}
\def\be{\beta}
\def\al{\alpha}
\def\ep{\epsilon}
\def\vep{\varepsilon}
\def\la{\lambda}	
\def\La{\Lambda}
\def\de{\delta}		
\def\De{\Delta}
\def\om{\omega}		
\def\Om{\Omega}
\def\sig{\sigma}	
\def\Sig{\Sigma}
\def\vphi{\varphi}

\def\CA{{\cal A}}	\def\CB{{\cal B}}	\def\CC{{\cal C}}
\def\CD{{\cal D}}	\def\CE{{\cal E}}	\def\CF{{\cal F}}
\def\CG{{\cal G}}	\def\CH{{\cal H}}	\def\CI{{\cal J}}
\def\CJ{{\cal J}}	\def\CK{{\cal K}}	\def\CL{{\cal L}}
\def\CM{{\cal M}}	\def\CN{{\cal N}}	\def\CO{{\cal O}}
\def\CP{{\cal P}}	\def\CQ{{\cal Q}}	\def\CR{{\cal R}}
\def\CS{{\cal S}}	\def\CT{{\cal T}}	\def\CU{{\cal U}}
\def\CV{{\cal V}}	\def\CW{{\cal W}}	\def\CX{{\cal X}}
\def\CY{{\cal Y}}	\def\CZ{{\cal Z}}


\def\barray{\begin{eqnarray}}
\def\earray{\end{eqnarray}}
\def\beq{\begin{equation}}
\def\eeq{\end{equation}}

\def\Tr{\rm Tr} 
\def\xvec{{\bf x}}
\def\kvec{{\bf k}}
\def\kvecp{{\bf k'}}
\def\omk{\om{\kvec}} 
\def\dk#1{\frac{d\kvec_{#1}}{(2\pi)^d}}
\def\2pid{(2\pi)^d}
\def\ket#1{|#1 \rangle}
\def\bra#1{\langle #1 |}
\def\vol{V}
\def\adag{a^\dagger}
\def\rme{{\rm e}}
\def\Im{{\rm Im}}
\def\pvec{{\bf p}}
\def\fermiS{\CS_F}
\def\cdag{c^\dagger}
\def\adag{a^\dagger}
\def\bdag{b^\dagger}
\def\vvec{{\bf v}}
\def\muhat{{\hat{\mu}}}
\def\vac{|0\rangle}
\def\pcut{{\Lambda_c}}
\def\chidot{\dot{\chi}}
\def\gradvec{\vec{\nabla}}
\def\psitilde{\tilde{\Psi}}
\def\psibar{\bar{\psi}}
\def\psidag{\psi^\dagger} 
\def\m{m_*}
\def\up{\uparrow}
\def\down{\downarrow}
\def\Qo{Q^{0}}
\def\vbar{\bar{v}}
\def\ubar{\bar{u}}
\def\smallhalf{{\textstyle \inv{2}}}
\def\smallsqrt{{\textstyle \inv{\sqrt{2}}}}
\def\rvec{{\bf r}}
\def\avec{{\bf a}}
\def\pivec{{\vec{\pi}}}
\def\svec{\vec{s}} 
\def\phivec{\vec{\phi}}
\def\daggerc{{\dagger_c}}
\def\Gfour{G^{(4)}}
\def\dim#1{\lbrack\!\lbrack #1 \rbrack\! \rbrack }
\def\qhat{{\hat{q}}}
\def\ghat{{\hat{g}}}
\def\nvec{{\vec{n}}}
\def\bull{$\bullet$}
\def\ghato{{\hat{g}_0}}
\def\r{r}
\def\deltaq{\delta_q}
\def\gcharge{g_q}
\def\gspin{g_s}
\def\deltas{\delta_s}
\def\gQC{g_{AF}} 
\def\ghatqc{\ghat_{AF}}
\def\xqc{x_{AF}}
\def\mhat{\hat{m}}
\def\xup{x_2}
\def\xdown{x_1}
\def\sigmavec{\vec{\sigma}}
\def\xopt{x_{\rm opt}}
\def\Lambdac{{\Lambda_c}}
\def\angstrom{{{\scriptstyle \circ} \atop A}     }
\def\AA{\leavevmode\setbox0=\hbox{h}\dimen0=\ht0 \advance\dimen0 by-1ex\rlap{
\raise.67\dimen0\hbox{\char'27}}A}
\def\ratio{\gamma}
\def\Phivec{{\vec{\Phi}}}
\def\singlet{\chi^- \chi^+} 
\def\mhat{{\hat{m}}}
\def\blue#1{{\color{blue}{#1}}}
\def\red#1{{\color{red}{#1}}}
\def\Im{{\rm Im}}
\def\Re{{\rm Re}}
\def\xstar{x_*}
\def\sech{{\rm sech}}
\def\Li{{\rm Li}}
\def\dim#1{{\rm dim}[#1]}
\def\ep{\epsilon}
\def\free{\CF}
\def\Fhat{\digamma}
\def\ftilde{\tilde{f}}
\def\muphys{\mu_{\rm phys}}
\def\xiprime{\tilde{\xi}}
\def\CI{\mathcal{I}}
\def\ko{k_0}
\def\Lambdastar{\Lambda_*}   
\def\gtilde{\tilde{g}} 
\def\ntilde{\tilde{n}}
\def\mutilde{\tilde{\mu}}

\section{I. Introduction}
The absence of conventional long range order in two-dimensional (2D) systems is well known to be a consequence of low energy fluctuations \cite{MerminWagner,Hohenberg}. Phase transitions at finite temperature in 2D are instead marked by a topolgical order as described by Berezinskii, Kosterlitz, and Thouless (BKT) \cite{Berezinskii,KT}. Quasi long range order is exhibited below the BKT transition temperature, where spatial correlations of the order parameter decay algebraically rather than exponentially. The destruction of this ordering, due to the unpairing of vortices, has been observed experimentally using atomic gases \cite{Hadzibabic}.

Ultracold atomic gases are well suited for the exploration of BKT theory, as they can be effectively constrained to 2D \cite{Martiyanov,Frohlich,Makhalov, Sommer, Tung, HaHung, Rammelmuller,Boettcher} using an optical lattice or harmonic trap, and because their interactions are highly tunable through the use of Feshbach resonances. Theoretical studies of BKT physics in two dimensions using quantum gases are numerous. Fermi gases have been used to explore Cooper pairing and superconductivity in 2D \cite{Randeria1, Randeria2}, as well as the normal Fermi liquid phase \cite{Engelbrecht,Engelbrecht2, Drut}. The superfluid transition and critical point of the 2D Fermi gas have also been treated \cite{Petrov2,Botelho,Miyake}. Bose Einstein condensation in 2D, particularly at weak coupling, has been widely studied \cite{Fisher, Holzmann}. The critical point of the 2D Bose gas has been obtained in this limit through classical $\phi^4$ theory \cite{Popov, Prokofev1,Prokofev2}, and with a renormalization group (RG) approach \cite{Rancon}. An RG analysis has also been used to examine the ground state of the Bose gas in arbitrary dimension \cite{Kolomeisky}. 

In this paper we apply the formalism developed in \cite{PyeTon} to study two-dimensional Bose and Fermi gases with both attractive and repulsive interactions. Inspired by the Yang-Yang equations of the Thermodynamical Bethe Ansatz \cite{YangYang}, the formalism is centered around an integral equation with a kernal based on the logarithm of the two-body S-matrix. The main advantage of our approach is that the two-dimensional integral equation admits approximate analytic solutions in terms of the Lambert $W$ function. Furthermore these approximate solutions remain useful beyond weak coupling, and as detailed in section III, they can be used to calculate any thermodynamic quantity of interest. In this work we primarily use this correspondence to calculate the critical points of the BKT transition for both Bose and Fermi gases. Below we summarize our chosen models, couplings, and formalism before describing our main results in sections IV and V.

\section{II. Renormalization Group, physical couplings,  and thermodynamic scaling functions}

The models considered in this paper are the simplest 
models of non-relativistic bosons and fermions with quartic
interactions.    The bosonic model is defined by the following action
for a complex scalar field $\phi$:
\beq
\label{bosonaction}
S =  \int d^2 \xvec \, dt \(  i \phi^\dagger  \d_t \phi - 
\frac{ |\vec{\nabla} \phi |^2}{2m}  - \frac{g}{4} (\phi^\dagger  \phi)^2 \).
\eeq
For fermions, due to the fermionic statistics, one needs at least a 
2-component field $\psi_{\up , \down} $:
\beq
\label{fermionaction}
S = \int d^2 \xvec  \, dt \(\sum_{\alpha=\up, \down}  
i \psi^\dagger_\alpha \d_t  \psi_\alpha  - 
\frac{|\vec{\nabla}  \psi_\alpha|^2}{2m}   - \frac{g}{2} 
\psi^\dagger_\up \psi_\up \psi^\dagger_\down \psi_\down \) .
\eeq
In both cases,  positive (negative) $g$ corresponds to 
repulsive (attractive) interactions.

Although classically  the coupling $g$ is dimensionless,    
in the quantum theory
it is not exactly marginal,  i.e.  it has a renormalization group  (RG) 
flow. One way to determine the beta function is to calculate the exact 2-body 
S-matrix \cite{PyeTon}.
Multi-loop Feynman diagrams are divergent,  and can be regularized with 
an ultraviolet cutoff $\Lambda$ in momentum space $\kvec$.
The result is
\beq
\label{2dfp.1}
\frac{ dg }{d \log \Lambda} = \frac{m g^2}{4\pi}.
\eeq
The above beta function is exact, i.e. there are no higher order 
corrections. It is characteristic of Berezinskii-Kosterlitz-Thouless 
transitions. For $g$ positive, it is marginally irrelevant in that at low energies 
$g$ flows to zero, whereas $g$ negative is marginally
relevant,  i.e. $g$ flows to $-\infty$.

There is a characteristic momentum scale in the problem, 
which plays the role of the physical coupling,
\beq
\label{lambdastar}
\Lambdastar  =   \Lambda \,  e^{4\pi/mg}.
\eeq
Since it is an RG invariant
\beq
\frac{ d \Lambdastar}{d \log \Lambda} = 0.
\eeq
The two-body S-matrix can be expressed entirely in terms of this scale:
\begin{align}
S( |\kvec| ) &= \frac{  4\pi /mg  + 
\log( 2 \Lambda/|\kvec|  ) - i\pi/2}
{4\pi /mg  + \log( 2 \Lambda/|\kvec| ) + i\pi/2} \label{2dS} \\
&= \frac{\log (2 \Lambdastar /|\kvec| ) - i \pi/2}{\log
 (2 \Lambdastar/|\kvec| ) + i \pi/2},
\end{align}
where $|\kvec| = |\kvec_1 - \kvec_2 |$ is the relative momentum
of the two incoming particles.  Note that because the two-body S-matrix can be calculated to all orders in the coupling, the second virial coefficient can be obtained exactly within our formalism, as explained in the appendix.  

We will define the physical scattering length $a_s$ as 
\beq
\label{as}
a_s = \inv{\Lambdastar}   =  \inv{\Lambda} \, e^{-4\pi/mg}.
\eeq
In two dimensions,  this choice is really just a convention.    
Comparing the scattering
amplitude based on the above S-matrix and the calculation 
in \cite{Rancon},   
the convention in the latter is  $a_s = e^{-\gamma_E} /  \Lambdastar$,  
where $\gamma_E $ is the Euler-Mascheroni  constant.

Thus as $g \to 0^+$,  i.e. approaches zero from above,  
the scattering length $a_s \to 0$.      
On the other hand $a_s \to \infty$  as  
$g\to 0^-$. 
Strong coupling,  $g\to \pm \infty$ just corresponds to a 
finite $a_s = 1/\Lambda$.  
Note that a negative scattering length is not physically possible, 
unlike in three dimensions.
For reference, the parameters in the weak and 
strong coupling limits are summarized in Table~\ref{table_limit}.
\begin{table}[h]
\caption{\label{table_limit}The parameter $\alpha$ is defined below in \eqref{variables}.  The weak coupling limit is obtained
with $\alpha \to \infty$ for repulsive interactions and $\alpha \to 0$
for attractive interactions. The strong coupling limit corresponds
to finite $\alpha$ in both cases. $\tilde{g}$, defined in \eqref{gtilde}, is also finite at strong coupling.}
\begin{ruledtabular}
\begin{tabular}{cccccccccccc}
 &\multicolumn{4}{c}{weak coupling}&\multicolumn{7}{c}{strong coupling} \\
\hline
 && $g$   & $a_s$ &$\alpha$&&$g$  & $a_s$ &$\alpha$
\\
repulsive && $0^+$ & 0&$\infty$&&$+\infty$ & $1/\Lambda^*$&$\mathcal{O}(1)$  \\
attractive &&$0^-$&$\infty$&$0^+$&&$-\infty$ & $1/\Lambda^*$&$\mathcal{O}(1)$  \\

\end{tabular}
\end{ruledtabular}
\end{table}

It is not physically meaningful to express properties in terms of $g$ since it is not a renormalization group invariant;   rather they should be 
expressed in terms of the scattering length $a_s$.    At finite temperature and density,  it is convenient to define the dimensionless variables:  
\beq
\label{variables}
\alpha =  \frac{\lambda_T}{a_s},  ~~~~~\mutilde = \frac{ \mu}{T}
\eeq
where $\lambda_T = \sqrt{2\pi/mT}$  is the thermal de Broglie wavelength and $\mu$ the chemical potential.      The scaled density
\beq
\label{scaleddensity}
\ntilde =  n \lambda_T^2  
\eeq is a function of $\mutilde$ and $\alpha$.     For the full equation of state involving pressure,  one also needs the free energy density: 
\beq
\label{freeenergy} 
\CF = - \frac{\pi^2}{6}  T   \lambda_T^{-2}  \, c(\mutilde, \alpha)
\eeq
where $c$ is defined below. With the above normalization,   $c=1$ for a free boson at zero chemical 
potential.    
In two dimensions the Fermi wave-vector is 
$k_F = \sqrt{2 \pi n} =  \lambda_T^{-1}\sqrt{2 \pi \ntilde}$
and the Fermi temperature $T_F = \pi n/m$,  which implies $T/T_F = 2/\ntilde$. In units of the Fermi energy, the energy per particle takes the form
\begin{equation}
\frac{E}{N} = \frac{\pi^2}{12}\left(\frac{T}{T_F}\right)^2 c(\tilde{\mu}, \alpha).
\label{epern}
\end{equation}
For comparison with  other calculations and experiments,   it is also useful to define a physical coupling $\gtilde$ as 
\beq
\label{gtilde}  
\gtilde = \frac{8 \pi} {| \log (n a_s^2) |}  = \frac{8 \pi} {| \log (\ntilde/\alpha^2)| } =\frac{4\pi}{\left|\log\(\frac{k_Fa_s}{\sqrt{2\pi}}\)\right|} 
\eeq
The above definition is such that at weak coupling $\gtilde \approx g$.

\section{III. Formalism}

In the formalism developed in \cite{PyeTon}, thermodynamic functions 
are determined by a pseudo-energy $\vep (\kvec )$  which represents  
a single particle energy in the presence of
interactions with all the other particles in the gas.
The occupation numbers take the  free particle form
\beq
f (\kvec) =   \inv{ e^{\beta \vep (\kvec) }  -s }, \qquad
n = \int \frac{d^2 \kvec}{(2\pi)^2} \,  f(\kvec)
\eeq
where $s=1$ ($s=-1$) corresponds to bosons (fermions).      
Consistent re-summation of 
all two particle interactions leads to an integral equation 
for $\vep (\kvec)$.
It is convenient to define the function 
\beq
\label{ydef}
y(\kvec) = e^{-(\vep (\kvec) -\omega_\kvec + \mu)/T}
\eeq
where $\omega_\kvec = \kvec^2/2m$.  Without interactions, $y=1$.     
With interactions  $y$ satisfies the non-linear integral equation
\beq
\label{yeq}
y (\kvec )  = 1  + \inv{T}   \int   
\frac{d^2 \kvec'}{(2\pi)^2}  \,  G_\pm (\kvec - \kvec') \,  
\frac{z}{e^{\omega_{\kvec'}  /T}  -s z y (\kvec')}
\eeq
where $z=e^{\mu/T}$,  
and  $G_{\pm}$ refers to bosons or fermions, respectively,  
with $G_- = G_+ /2$.
Henceforth  $G$ without a subscript will refer to $G_+$.    
The fermionic case will be studied in section V. 

The kernel $G$ is related to the logarithm of the two-body S-matrix,
\begin{align}
\label{Gdef}
G (\kvec - \kvec' )  &= -\frac{4i}{m} \log S(\kvec-\kvec')\nonumber \\ 
&=   \frac{8}{m}  {\rm arccot}  \[\frac{2}{\pi} \log \(   a_s |\kvec - \kvec'| /2 \)   \].
\end{align}
Rescaling $\kvec  \to \sqrt{2 m T} \, \kvec$,  the scaling functions are 
then given by the following integrals ($k = |\kvec|$):
\begin{align}
\label{qint}
\ntilde &= 2 \int_0^\infty  dk \,  k    \frac{y(k) z}{e^{k^2}  - s  y(k) z}, \\
\label{cint}
c &=  -  \frac{12}{\pi^2}   \int_0^\infty  dk \,  k  
\[ s \log \( 1 - s z y(k) e^{-k^2} \) + 
\inv{2}  \frac{z (y(k) -1)}{e^{k^2} - s z y(k)} \].
\end{align}
The integral equation \eqref{yeq} then becomes
\beq
\label{yeq2}
y (k )  = 1  + \frac{m}{2\pi^2}   \int   d^2 \kvec'  \,  
G  (\kvec - \kvec') \,  
\frac{z}{e^{k'^2}  -s z y (k')},
\eeq
where the kernel is 
\beq
\label{Gdef2}
G (\kvec - \kvec' ) =  \frac{8}{m}  {\rm arccot}  \[  
\frac{2}{\pi} \log \(  \frac{ \sqrt{\pi}  |\kvec - \kvec'|}{\alpha}  \)   \].
\eeq

The free theory limit corresponds to  \emph{both},  
$\alpha \to \infty$  ($g\to 0^+$)  and $\alpha \to 0$ ($g \to 0^-$). 
It will be convenient to write the integral equation \eqref{yeq2} as
\beq
\label{Hint} 
y(k)  =   1 + \int_0^\infty  dk'  \, k' \,  H(k,k') \,  
\frac{z}{e^{k'^2} - sz\, y(k') },
\eeq
where 
\begin{widetext}
\beq
\label{Hdef}
H(k,k')  = \frac{4}{\pi^2}  \int_0^{2\pi}  d\theta  \, {\rm arccot}  \[
\inv{\pi}  
\log \(  \frac{ \pi (k^2 + k'^2 -2 k k' \cos \theta)}{\alpha^2} \) \].
\eeq
\end{widetext}

In theory the integral equation \eqref{yeq} is valid for all interaction strengths and temperatures, including $T=0$. However, as our formalism only includes a pseudo-energy for single particles, the integral equation is only solvable in the absence of bound states, limiting its applicability at zero temperature. Low temperature results are accurate only if the gas remains in the normal phase. As a concrete example, we calculated Tan's contact for the attractive Fermi gas in the normal phase for temperatures as low as $T/T_F=0.001$ (see Figure 5 and the discussion in Section V).

 We also expect solutions of the integral equation to become less reliable if the coupling is large enough to result in appreciable many-body interactions, as only two-body effects are considered. If we take as a measure of strong verses weak coupling the constant 
$\gtilde$ defined in \eqref{gtilde},   then because of the logarithm,   
 $\gtilde \approx 1$ requires (for repulsive interactions) a very large $\alpha$,  on the order of $10^6$ 
 (see below).      For such large $\alpha$,   in practice we can disregard the
 momentum dependence of the above kernel.    We checked numerically 
 that retaining the $k$ dependence did not substantially alter our results below.
 Thus,  henceforth 
 \beq
\label{Hweak}
H \approx  \frac{8}{\pi}  {\rm arccot} \[ -\frac{2}{\pi} \log \alpha  \]   
\approx 
- \frac{4}{\log \alpha}  \to  0^{\pm}.
\eeq
is a constant that differs in sign depending on 
whether the interaction is attractive or repulsive.

\section{IV. Bosons}

Since the kernel is a constant,  $y$ is also a constant.   
The integral over $k$ can be
performed and the integral equation  \eqref{Hint} becomes the 
transcendental equation
\beq
\label{ytrans} 
y (y-1) =   \frac{ 2 \log (1-yz)}{ \log \alpha} .
\eeq
The equation of state is then 
\beq
\label{eqstate}
\ntilde =   - \log(1-yz) .
\eeq
Given  the solution to \eqref{ytrans} for $y$ as a function of $z$,  
the above equation determines how the density depends on the 
chemical potential and temperature.

For very weak coupling,  $y \approx 1$,  and \eqref{ytrans} can be approximated as  
\beq
\label{ytrans2} 
y  = 1+   \frac{ 2\log (1-yz)}{ \log \alpha} 
\eeq
which can be expressed in terms of the Lambert $W$-function. The $W$ function 
by definition satisfies 
\beq
\label{LambertDef}
W(u) e^{W(u)} = u.
\eeq
One then has the following solution of the transcendental equation:
\beq
\label{zeq}
y = 1 - a \log (1- y/b)\quad \Rightarrow \quad 
y = b + a W \( - \frac{b}{a}  e^{(1-b)/a} \).
\eeq
Thus 
\beq
\label{YLambert}
y =  \inv{z}  - 
\frac{2}{\log \alpha} \, W\( \frac{\log \alpha }{2z} 
e^{ (1/z -1) \log \sqrt{\alpha} } \). 
\eeq

For real $u$, $W(u)$ has two real valued branches, as shown in Figure \ref{Wbranch}  
below.    If $u\ge0$
there is only the principal branch denoted by $W_0(u)$, and
if $-1/e \le u < 0$ we have the principal branch $W_0$ and also
the secondary branch $W_{-1}(u)$.   The two branches only coincide when $u=-1/e$, 
where $W_0(-1/e)=W_{-1}(-1/e)=-1$.  
\subsection{A. Repulsive Bosons}

    For the repulsive case the argument of $W$ is positive
and one should chose the principle branch,   henceforth simply denoted as $W$.   
\begin{figure}[h]
\centering
\includegraphics[scale=0.6]{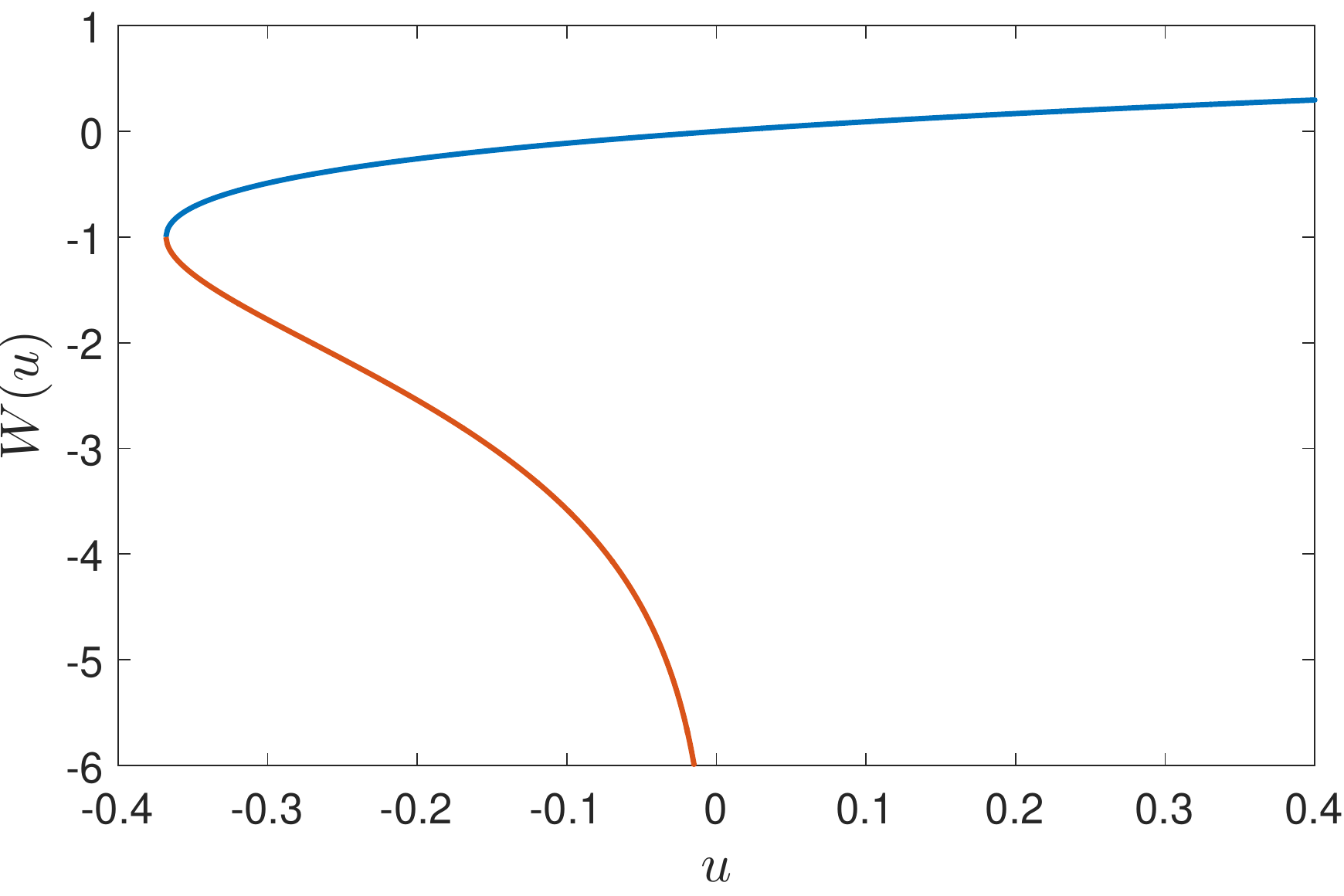}
\caption{(color online) The two real branches of the Lambert $W$-function.  The top blue part is the principal branch $W_0 (u) $,  whereas the lower orange part
is $W_{-1} (u)$. The branches meet at $u=-1/e$. }
\label{Wbranch}
\end{figure}

\begin{figure}[h]
\centering
\includegraphics[scale=0.6]{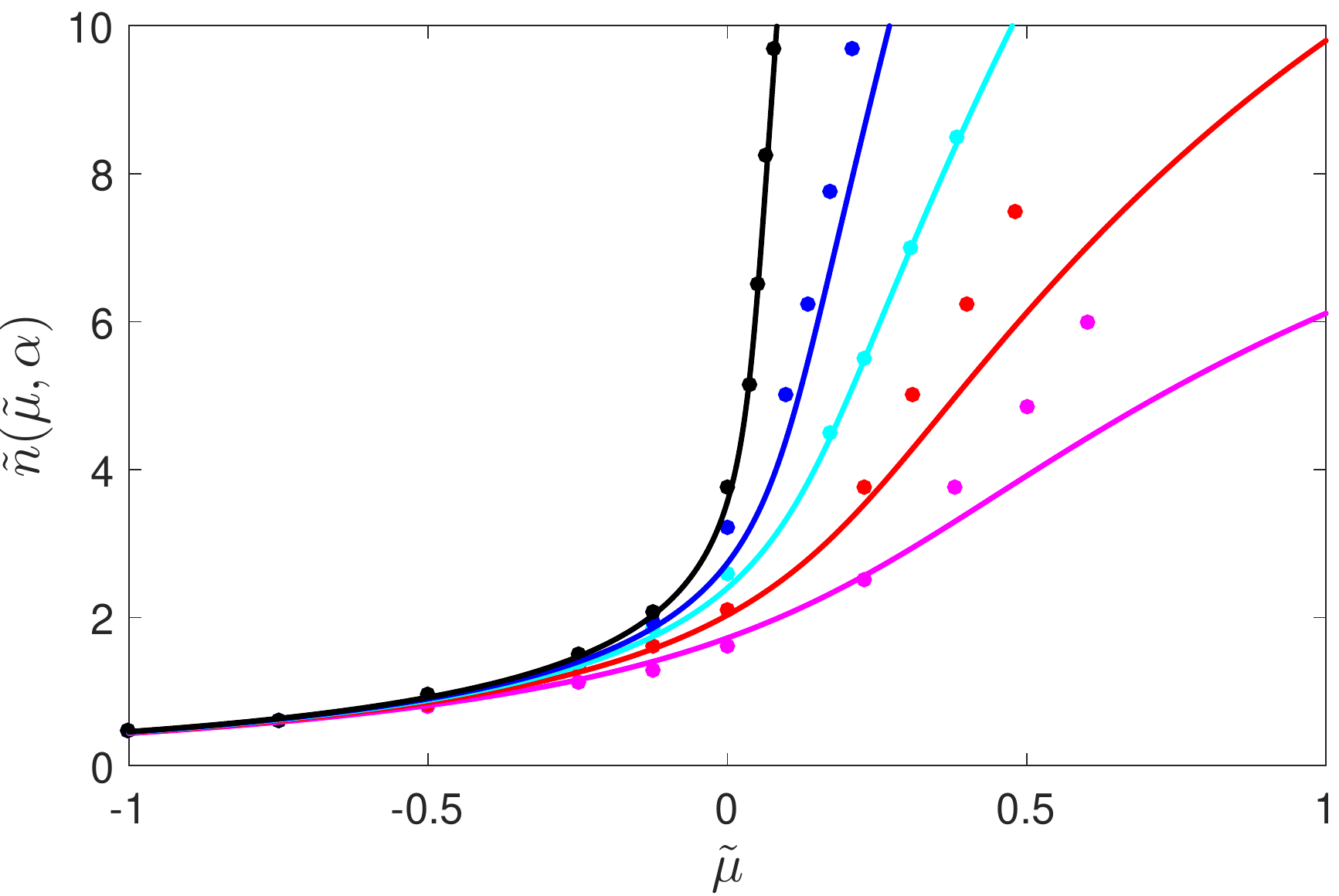}
\caption{(color online) The scaled density $\ntilde$ as a function of the scaled chemical potential $\mutilde$ for various 
$\gtilde_0$. From top to bottom, the solid colored lines correspond to $\gtilde_0 = 0.05, 0.15, 0.24, 0.41, 0.66$ respectively. The colored points, which have the same ordering from top to bottom, are experimental data estimated from \cite{HaHung}.}
\label{eqState}
\end{figure}

In order to compare with other theories and experiments,   we wish to plot $\ntilde$  as a function of 
$\mutilde$  for various $\gtilde$.    The above equations give explicit expressions in terms of $\alpha$ rather than
$\gtilde$.   However the primary variation of $\gtilde$ comes from the variation of $\alpha$.    Therefore we plot
$\ntilde$ as a function $\mutilde$  for a fixed  $\alpha$.   Along such a curve $\gtilde$  is nearly constant, thus it is meaningful to associate each fixed-$\alpha$ curve with $\gtilde_0   =   \gtilde (\mutilde = 0^- , \alpha)$.   
Our results, which use the approximation \eqref{YLambert}, are compared to experimental data in Figure \ref{eqState} for $\gtilde_0$  ranging  between $0.05$ and $0.66$. 
Due to the logarithms,  this requires a very large range of $\alpha$.   For instance  $\gtilde_0=0.05$ corresponds to 
$\alpha \approx 10^{110}$ whereas $\gtilde_0 = 0.66$ coincides with $\alpha \approx 10^8$.      

As $\mutilde \to - \infty$ the behavior is of a free gas
$\ntilde \approx - \log (1-z)$.    There are significant downshifts at finite $\mutilde$ which increase with $\gtilde$.   
These results compare reasonably well  with the measurements in \cite{HaHung}.

\def\sqalpha{\sqrt{\alpha}}
\def\mutilde{ \tilde{\mu}}

At some critical density $\ntilde_c$  the gas is known to become  a superfluid.     In order to motivate our analysis of
the critical point,  let us consider ordinary BEC of free particles in three spatial dimensions.      Here the kernel $G = 0$,  
and $y(\kvec) = 1$,    which implies $\vep (\kvec) = \omega (\kvec) - \mu$.    At the critical point  the occupation
number $f(\kvec = 0)$ diverges,  implying $\vep (0) = \mu_c = 0$.    This property is reflected in the analytic properties of
the density as a function of $z$ as follows.    One has $\ntilde =  n \lambda_T^3 =  {\rm Li}_{3/2} (z)$  where ${\rm Li}$ is the polylogarithm.   
The latter  has a branch cut along the real axis $\Re (z) > 1$ where the density $\ntilde$ develops an imaginary part.    Therefore the 
critical point is $z_c = 1$  and  $\ntilde_c =  {\rm Li}_{3/2}  (z_c) = \zeta (3/2)$  where $\zeta$ is the Riemann zeta function.    In two dimensions
this leads to $\zeta (1)$ which diverges due to the pole of $\zeta$ at $z=1$.   

For a two dimensional interacting gas, as in the 3D non-interacting case,  at the critical point the scaled density $\ntilde$  develops an imaginary part. This occurs 
for $y \approx 1/z$ where  the RHS of \eqref{ytrans} has a branch cut.  
 In order to study this analytically using known functions,
we consider the approximate solution of \eqref{ytrans} given by \eqref{YLambert}. Taking $y=1/z_c$ implies
\beq
\label{zc1} 
W\( \frac{\log \sqrt{\alpha} }{z_c} 
e^{ (1/z_c -1) \log \sqrt{\alpha} } \) = 0.
\eeq
    Since the argument of $W$ is arbitrarily large and positive for weakly coupled repulsive interactions, the approximation $W_0(u) \approx \log u$ can be used, giving
\beq
\label{zc2}  
\log \( \frac{ \log \sqrt{\alpha}}{z_c} \) + \log \sqrt{\alpha} \left(\frac{1}{z_c}-1\right) = 0.
\eeq
The solution $z_c$ to the above equation can again be  expressed in terms of the Lambert $W$ function 
\beq
\label{zc3}  
z_c = \frac{\log \sqrt{\alpha}}{W\left(\sqrt{\alpha}\right)}.
\eeq
Noting to second order, for large $u$, $W_{0} (u)  \approx \log u - \log \log u$, one can use the above equation and \eqref{eqstate}  to  compute the critical chemical potential and density:
\barray
\nonumber 
\mutilde_c  &\approx&  \frac{  \log \log \sqalpha }{\log \sqalpha}  \\
\label{munc} 
\ntilde_c  & \approx &  \log \log \sqalpha.
\earray

We now compare  \eqref{munc}  with known results and experiments. From the scattering length definition given by \eqref{as} we see
$\log \sqalpha \approx 2 \pi  / mg  $. This leads to
\barray
\nonumber 
\mutilde_c  &\approx&   \frac{mg}{2\pi}  \log  \(   \frac{ 2 \xi_\mu }{mg}  \)  \\
\label{munc2} 
\ntilde_c  & \approx &  \log \(  \frac{ 2 \xi}{mg} \)  
\earray
with  $\xi = \xi_\mu = \pi$.      The above functional dependence on $g$ agrees with \cite{Prokofev1, HaHung}    except for the constants inside the logarithm,  
where it was found that $\xi_\mu \approx 13.2$ and $\xi \approx 380.$  
We do not understand the reason for this discrepancy.   The simplest explanation is that
we are neglecting intrinsic 3-body interactions and higher;   however it is hard to see how these would lead to the same functional form as in \eqref{munc2} with just modifications of the $\xi$'s.    
It seems more likely to be  an effect of our simplifications of the kernel in the integral equation,   which effectively neglected its momentum dependence in the limit of large 
$\alpha$.       Although 
we did not find any significant evidence of this numerically,   it remains possible that we did not 
treat the integral equation properly in the infra-red,  i.e. low $k$.

\subsection{B. Attractive Bosons}

\begin{figure}[h]
\centering
\includegraphics[scale=0.6]{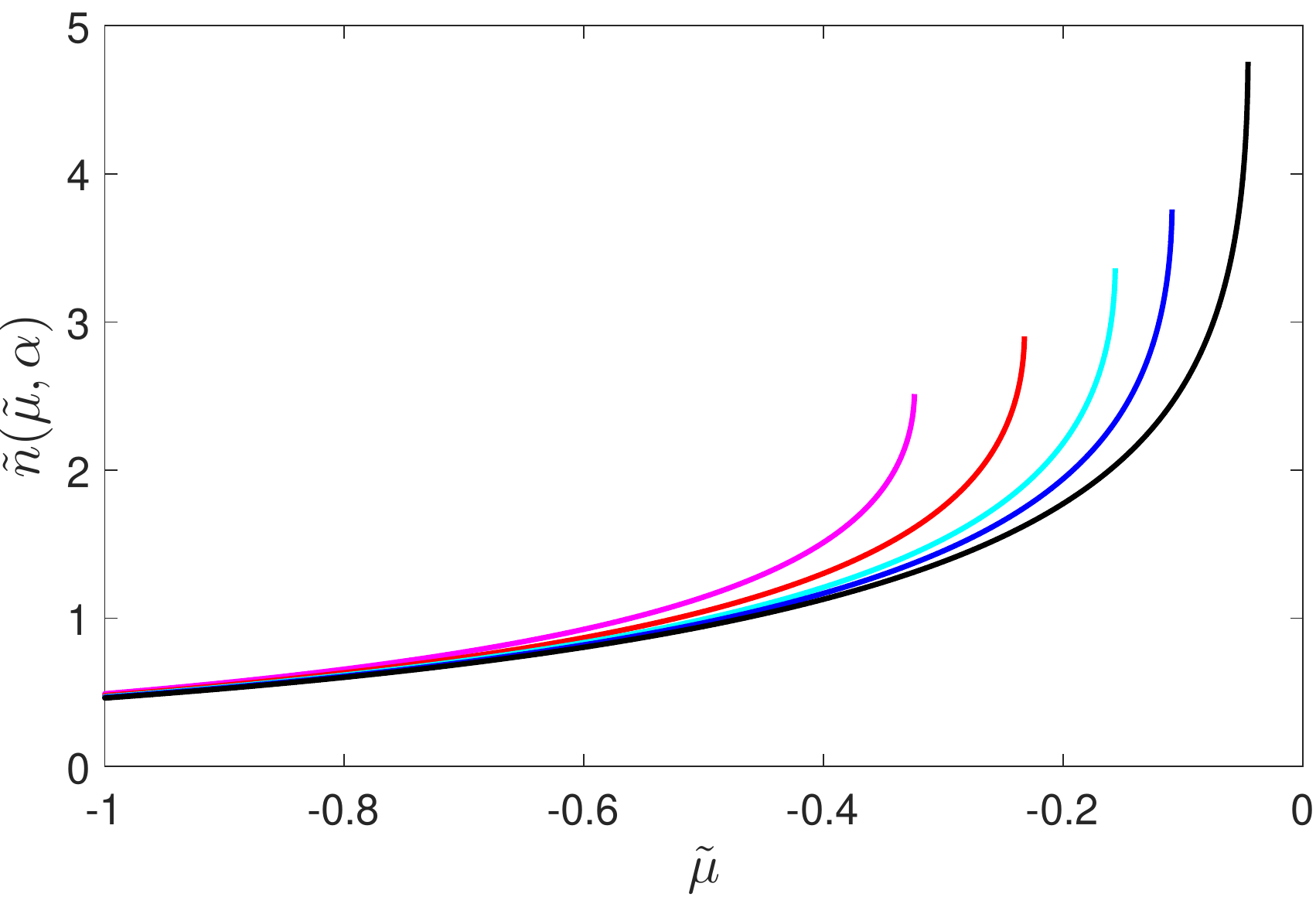}
\caption{(color online) The scaled density $\ntilde$ as a function of the scaled chemical potential $\mutilde$  for various 
$\gtilde_0$. From bottom to top, the solid colored lines correspond to $\gtilde_0 = 0.05, 0.15, 0.24, 0.41$, and $0.66$ respectively. Each $\tilde{n}$ curve terminates at a specific chemical potential, after which no solution to \eqref{ytrans2} exists for larger values of $\tilde{\mu}$.}\label{attracteqState}
\end{figure}

For attractive interactions in the weak coupling limit, \eqref{Hweak} is positive, and $\alpha \rightarrow 0^+$. In this regime there exists only very small regions of parameter space where there is a solution to 
\eqref{ytrans}. This instability is reflected in the equation of state, as shown in Figure \ref{attracteqState}.

 At the critical point, $\ntilde$ again becomes complex. The main difference in repeating the analysis of the previous section is now, for attractive interactions, the argument of $W$ in \eqref{YLambert} is arbitrarily small and negative and one must choose the secondary branch rather than the principle branch. After setting $y=1/z_c$ we find
 \beq
\label{Attractzc1} 
W_{-1}\( \frac{\log \sqrt{\alpha} }{z_c} 
e^{ (1/z_c -1) \log \sqrt{\alpha} } \) = 0.
\eeq
    Utilizing the asymptotic expansion of the secondary branch $W_{-1}(-\frac{1}{u}) \approx -\log u-\log \log u$ gives
\beq
\label{Attractzc2}  
\log \(\frac{ \lvert \log \sqrt{\alpha}\rvert}{z_c}\) - \left(\frac{1}{z_c}-1\right)\lvert \log \sqrt{\alpha}\rvert = 0,
\eeq
from which it follows
\beq
\label{Attractzc3}  
z_c = \frac{\log \sqrt{\alpha}}{W_{-1}\(-\sqalpha \)}.
\eeq
As in the repulsive case the critical density can be computed by inserting the above expression into \eqref{eqstate}. Using the asymptotic expansion of $W_{-1}$ a second time results in
\barray
\nonumber 
\mutilde_c  &\approx&  \frac{  -\log \lvert \log \sqalpha \rvert }{ \lvert \log \sqalpha \rvert}  \\
\label{Attractmunc} 
\ntilde_c  & \approx &  \log  \lvert \log \sqalpha \rvert.
\earray

\section{V. Fermions} \label{sec:fermions}
In the fermionic case the kernal $G_- = G_+/2$. The analogs of \eqref{ytrans} and \eqref{eqstate} for fermions at weak coupling are then
\begin{equation}
y(y-1) = -\frac{\log(1+yz)}{\log \alpha}
\label{fermi_ytrans}
\end{equation} and
\begin{equation}
\tilde{n}=\log(1+yz).
\label{fermi_eqstate}
\end{equation} Mirroring our treatment of bosons, we find approximate solutions for $y$ in the attractive ($\alpha \rightarrow 0$)
\beq
\label{atfermi_YLambert}
y =  -\inv{z}  + 
\frac{1}{\log \alpha} \, W_{-1}\( \frac{\log \alpha }{z} 
e^{ (1/z -1) \log \sqrt{\alpha} } \) 
\eeq
and repulsive ($\alpha \rightarrow \infty$)
\beq
\label{repfermi_YLambert}
y =  -\inv{z}  + 
\frac{1}{\log \alpha} \, W_0\( \frac{\log \alpha }{z} 
e^{ (1/z -1) \log \sqrt{\alpha} } \) 
\eeq
regimes in terms of the Lambert function.

\section*{A. Attractive Fermions}
A useful quantity to calculate is the contact parameter, $C$, which is set by the antiparallel spin pair correlation function $g_{\uparrow \downarrow} (r)$ at  short  distances ($r \ll 1/k_F$). Tan's relations provide a connection between  $C$, and thus the short range interactions of the system, and macroscopic quantities such as the pressure of the gas \cite{Tan}. One can define a dimensionless contact in terms of a derivative of the energy with respect to the interaction parameter
\begin{equation}
C'=C/k_F^2=\pi \frac{ d \frac{E}{E_F}}{d\log(k_Fa_s)}.
\end{equation} Note since $T/T_F=2/\tilde{n}$ this derivative can be calculated explicitly with our formalism. Using the approximations given by \eqref{fermi_eqstate} and \eqref{atfermi_YLambert}:
\begin{equation}
C'=\frac{2\pi\left[\log \alpha^{(1+	1/z)}-W_{-1}(v)\right]}{ \log \left(\frac{zW_{-1}(v)}{\log \alpha}\right)^2\left[ \log \alpha\left(W_{-1}(v)+1 \right)\right]}
\label{cprime}
\end{equation} where $v=\alpha^{(1+1/z)}\log \alpha^{1/z}$.

\begin{figure}[h]
\centering
\includegraphics[scale=0.6]{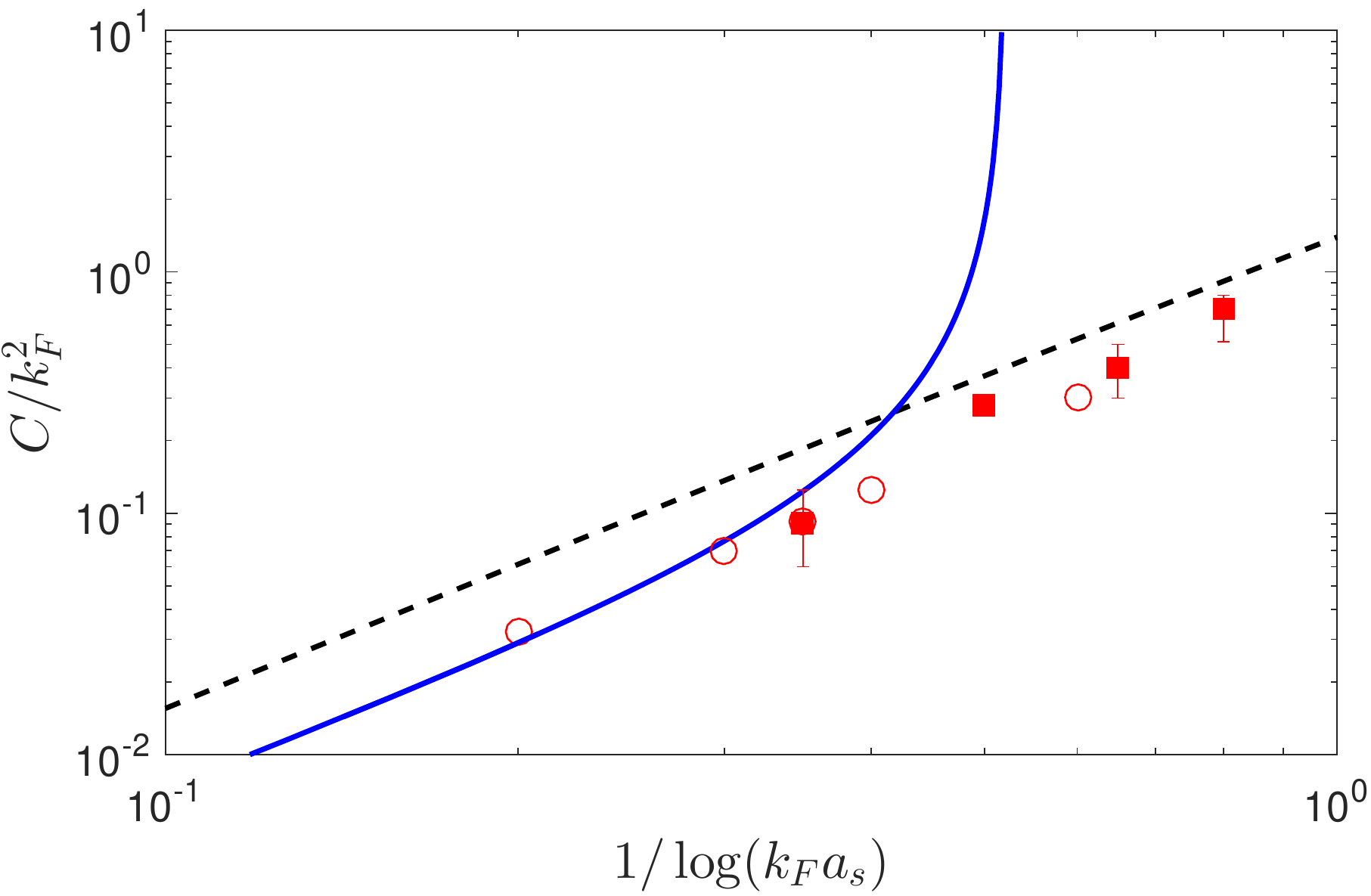}
\caption{Dimensionless contact $C'$ vs. $1/\log(k_Fa_s)$ at $T/T_F=0.27$. The red squares and error bars are experimental data estimated from \cite{Frohlich}. The red open circles are from the same group, but calculated from a model which accounts for their specific experimental configuration. The dashed black line is a 2nd order $T=0$ homogenous Fermi liquid theory result, and the solid blue curve uses the Lambert approximation \eqref{atfermi_YLambert} to calculate $C'$.}\label{attractive_contact}
\end{figure}

For the 2D attractive fermi gas, the contact was recently measured experimentally at $T/T_F=0.27$ \cite{Frohlich}. In Figure \ref{attractive_contact} below we compare $C'$ as calculated with our approximations to experiment, as well as to a $T=0$ Fermi liquid theory result \cite{Engelbrecht2}. Our $C'$ compares favorably with the experimental measurements until diverging abruptly as $\alpha \rightarrow 1$.

\begin{figure}[h]
\centering
\includegraphics[scale=0.6]{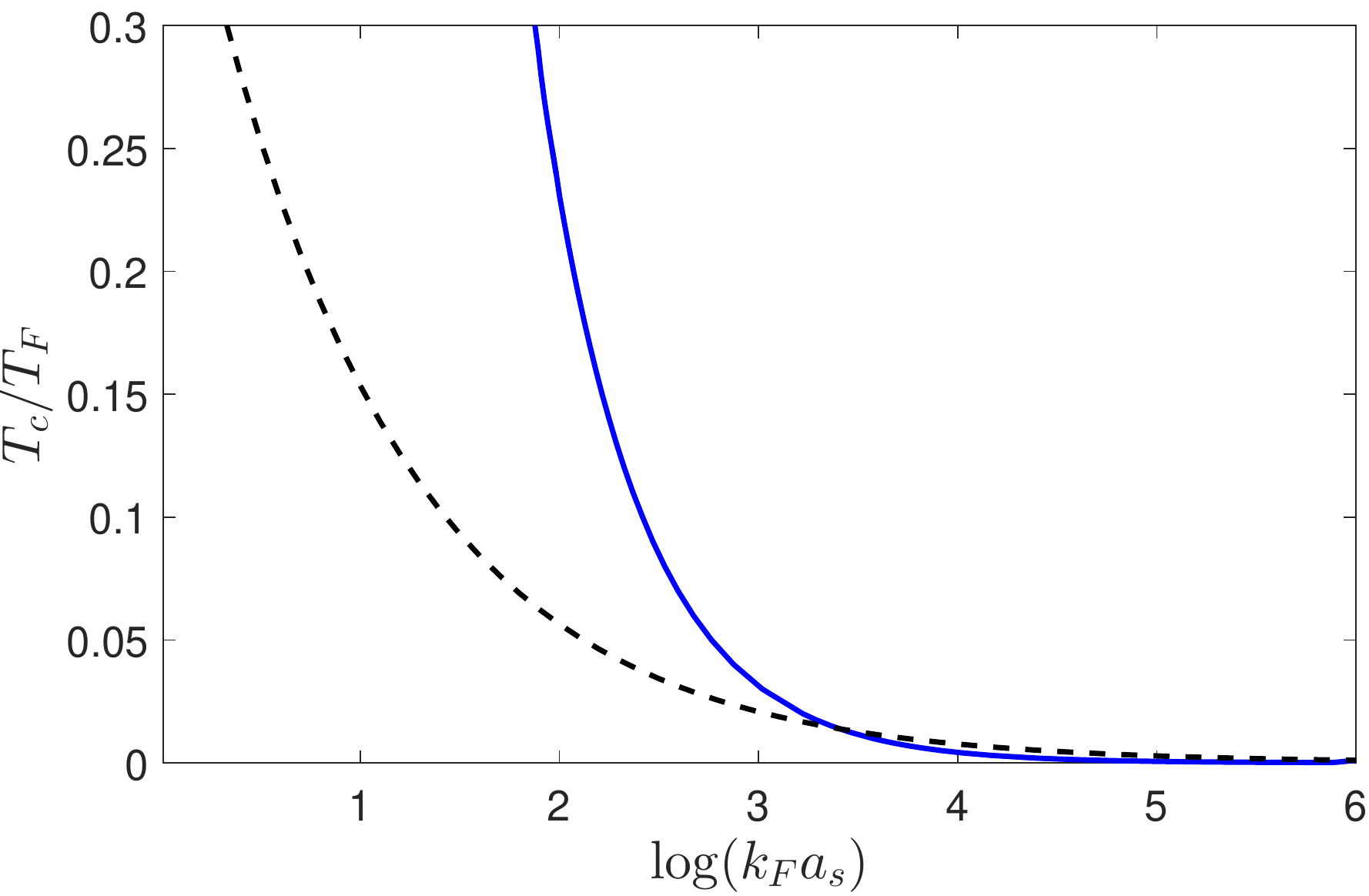}
\caption{$T_c/T_F$ as a function of the interaction parameter $\log(k_Fa_s)$. The dashed black line is the mean field theory result \eqref{att_mft} and the blue is obtained through the method described in the text. Up to $\log(k_Fa_s)\approx 3$ our $T_c/T_F$ closely matches the mean field theory prediction.}\label{att_fermiphase}
\end{figure}

Since $C'$ is proportional to the number of atomic pairs, which follows from the relation between the contact and $g_{\uparrow \downarrow} (r)$ \cite{Werner}, a divergence in $C'$ may signal a phase transition. Identifying the critical point of the BKT transition with a diverging $C'$ yields the phase diagram shown in Figure \ref{att_fermiphase}.

In the 2D BCS limit $\log \left( k_Fa_s\right) \gg 1$, the critical temperature of the superfluid transition has been calculated using mean field theory \cite{Petrov2,Miyake}
\begin{equation}
\frac{T_c}{T_F}=\frac{2e^{\gamma-1}}{\pi k_Fa_s}=\frac{c_{MF}}{k_Fa_s}.
\label{att_mft}
\end{equation} Fitting our phase boundary to a second order model
\begin{equation}
\frac{T_c}{T_F}=\frac{c_1}{k_Fa_s}+\frac{c_2}{(k_Fa_s)^2}
\end{equation} gives $c_1=0.865=2.08c_{MF}$ and $c_2=6.07$. Our results begin to significantly depart from those of mean field theory around $\log\left(k_Fa_s\right)=3$ which corresponds to $\tilde{g} \approx 6$. This is well into the regime of strong interactions, so it's unsurprising we deviate from a mean field theory prediction.

Before considering repulsive interactions, we note that Monte Carlo simulations have been used to calculate both the contact and energy per particle at $T=0$ on both sides of the BEC-BCS crossover. For example, in \cite{Bertaina} for $\log(k_Fa_s) = 5.18$, the normalized energy per particle is reported as $E/N=0.821$. To compare to this data we calculate $E/N$ using \eqref{epern} at low temperature. For convenience we take $T/T_F=0.01$, which fixes the density $\tilde{n}$. Since the interaction parameter $\log (k_F a_{s})$ only depends on $\tilde{n}(\tilde{\mu},\alpha)$ and $\alpha$, setting $\log(k_Fa_s) = 5.18$ uniquely determines $\alpha$. With $\alpha$ and $\tilde{n}(\tilde{\mu},\alpha)$ known, a corresponding $\tilde{\mu}$ can be found numerically, and used to calculate the energy per particle. Utilizing the approximations \eqref{fermi_eqstate} and \eqref{atfermi_YLambert} throughout, we obtain $E/N = 0.995$. Although $E/N$ is provided for a range of interaction strengths, most of the data presented in \cite{Bertaina} occurs in the presence of a bound state, which our formalism is not suited to handle.

\section*{B. Repulsive Fermions}
While we don't have experimental data to compare to on the BEC side, the contact is equivalent to \eqref{cprime} with $W(v)$ in place of $W_{-1}(v)$.  As in the attractive case, a sharp increase in $C'$ is observed (see Figure 6) which we take as indication of a phase transition. 
In the BEC limit of $\log \left( k_F a_s \right) \ll -1$, the predicted critical temperature is
\begin{equation}
\frac{T_c}{T_F}=\frac{1}{2}\left(\log \left(\frac{\xi}{2\pi}\log \left( \frac{2\sqrt{\pi}}{k_Fa_s}\right)\right)\right)^{-1}
\label{rep_mft}
\end{equation} with $\xi = 380$ \cite{Prokofev1, Petrov2}. We instead find $T_c/T_F$ to be more consistent with the value $\xi=\pi$ from \eqref{munc2}, as shown in Figure 7. The return of this discrepancy is expected based on the analysis in section IV, as in this limit the system behaves as a weak Bose gas.
\begin{figure}[h]
\centering
\includegraphics[scale=0.6]{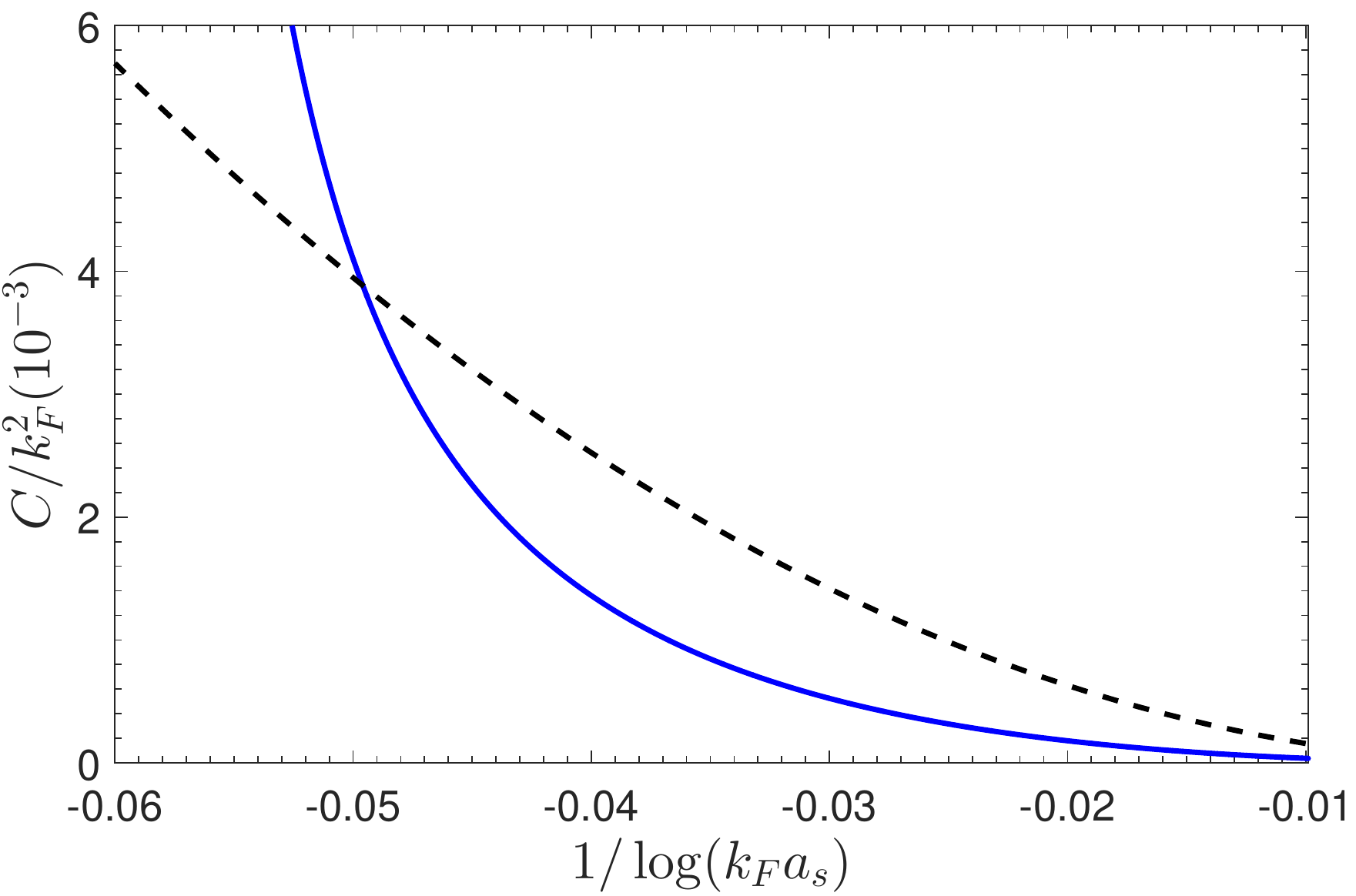}
\caption{Dimensionless contact $C'$ vs. $1/\log(k_Fa_s)$ at $T/T_F=0.1$. The dashed black line is the 2nd order $T=0$ homogenous Fermi liquid theory prediction, and the blue uses the Lambert approximation \eqref{repfermi_YLambert} to calculate $C'$.}\label{repulsive_contact}
\end{figure}

\begin{figure}[h]
\centering
\includegraphics[scale=0.6]{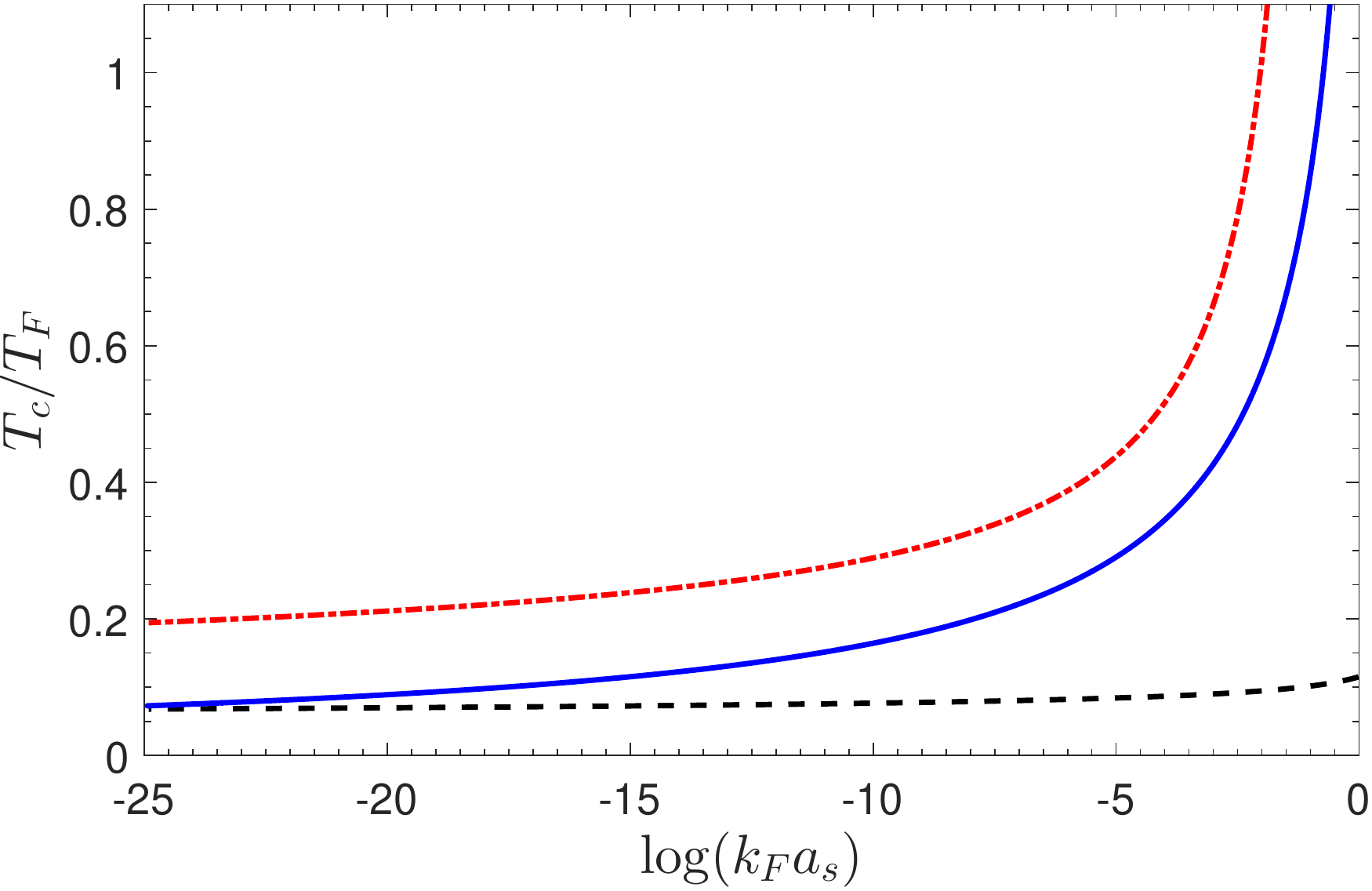}
\caption{$T_c/T_F$ as a function of the interaction parameter $\log(k_Fa_s)$.The dashed black line is the mean field theory result given by \eqref{rep_mft}. The red dash-dot line uses the same model except with our previously calculated $\xi= \pi$ instead of $\xi=380$. The solid blue curve is determined by the behavior of $C'$, as described in the text.}\label{rep_fermiphase}
\end{figure}

\section{VI. Conclusions}
The S-matrix-based formalism developed in \cite{PyeTon} has been applied to two-dimensional Bose and Fermi gases. The main obstacle in utilizing this method to extract measurable thermodynamic functions is solving an integral equation whose kernal takes a particularly complicated form in two dimensions. This makes exact solution of the integral equation an impossible task, and numerical treatments computationally intensive. Fortunately, in the limits $\alpha \ll 1$ and $\alpha \gg 1$ the momentum dependence of the kernal becomes irrelevant, and elegant solutions of the integral equation can be written in terms of the Lambert W function.

While we initially anticipated these approximate solutions would only remain valid in the limit of extremely weak coupling, it turns out $\tilde{g}$ is a more appropriate measure of coupling strength than $\alpha$, due to its explicit density dependence. For strong coupling, $\tilde{g} \gtrsim 1$ which is reached in the $\alpha \ll 1$ or $\alpha \gg 1$ limit depending on the sign of the interaction. 

For bosons, we were able to recover the well-established logarithmic functional form of the critical density and chemical potential with the Lambert approximation, up to a constant obtained with Monte Carlo methods \cite{Prokofev1}. For fermions our approximations result in an explicit expression for the contact parameter, from which the critical temperature of the BKT transition has been deduced.  This novel approach agrees with known weak coupling mean field theory calculations \cite{Miyake,Petrov2}.

As two-dimensional gases are poised to garner even greater attention in the near future, we hope our explicit analytic results 
 are found to be useful.

\section{Acknowledgments}

We thank John Stout for collaboration in the early stages of this work which led to the Appendix. 

\section{Appendix: Virial expansion}

\def\dk#1{\frac{d^2 #1}{(2 \pi)^2}} 

We did not use the following results in the body of the article,  however we present them here since they may be useful
in future studies.

The virial expansion is  formally  defined as a series expansion of $\CF$ in powers of  the fugacity $z$:
\barray
\nonumber 
- \CF  \lambda_T^2 /T &=&  \sum_{n=1}^\infty   b_n z^n  \\
\label{virial.1}
n \lambda_T^2  &=&  \sum_{n=1}^\infty  n \, b_n  z^n  
\earray
where the second relation follows from $n = - \d \CF/ \d \mu$.  
  In the free theory,  the series expansion 
of the poly-logarithm  $s \Li_{2} (s z )$  gives  
$b_n =  s^{n+1} / n^2 $.   

As explained in \cite{Roditi}
our formalism gives the corrections  to $b_2$
and $b_3 $:
\begin{align}
\nonumber
 b_2   =  \frac{s}{4}  +   \frac{\lambda_T^2}{2T}  \int & \dk{\kvec}  \dk{\kvec'} \\ \times & \left(e^{-\omega_\kvec /T}  e^{- \omega_{\kvec' }/T}  G_\pm (\kvec - \kvec')\right),
 \label{bas} 
\end{align}
\begin{align}
\nonumber 
b_3   = \inv{9} +   \frac{s \lambda_T^2}{2T}  \int   \dk{\kvec} &  \dk{\kvec'} e^{-\omega_\kvec /T}  e^{- \omega_\kvec' /T}  G_\pm (\kvec - \kvec')  \\ \times & 
\( e^{-\omega_\kvec /T}  + e^{-\omega_{\kvec'} /T}  \).  \nonumber
\end{align}

\begin{figure}
\centering
\includegraphics[scale=0.6]{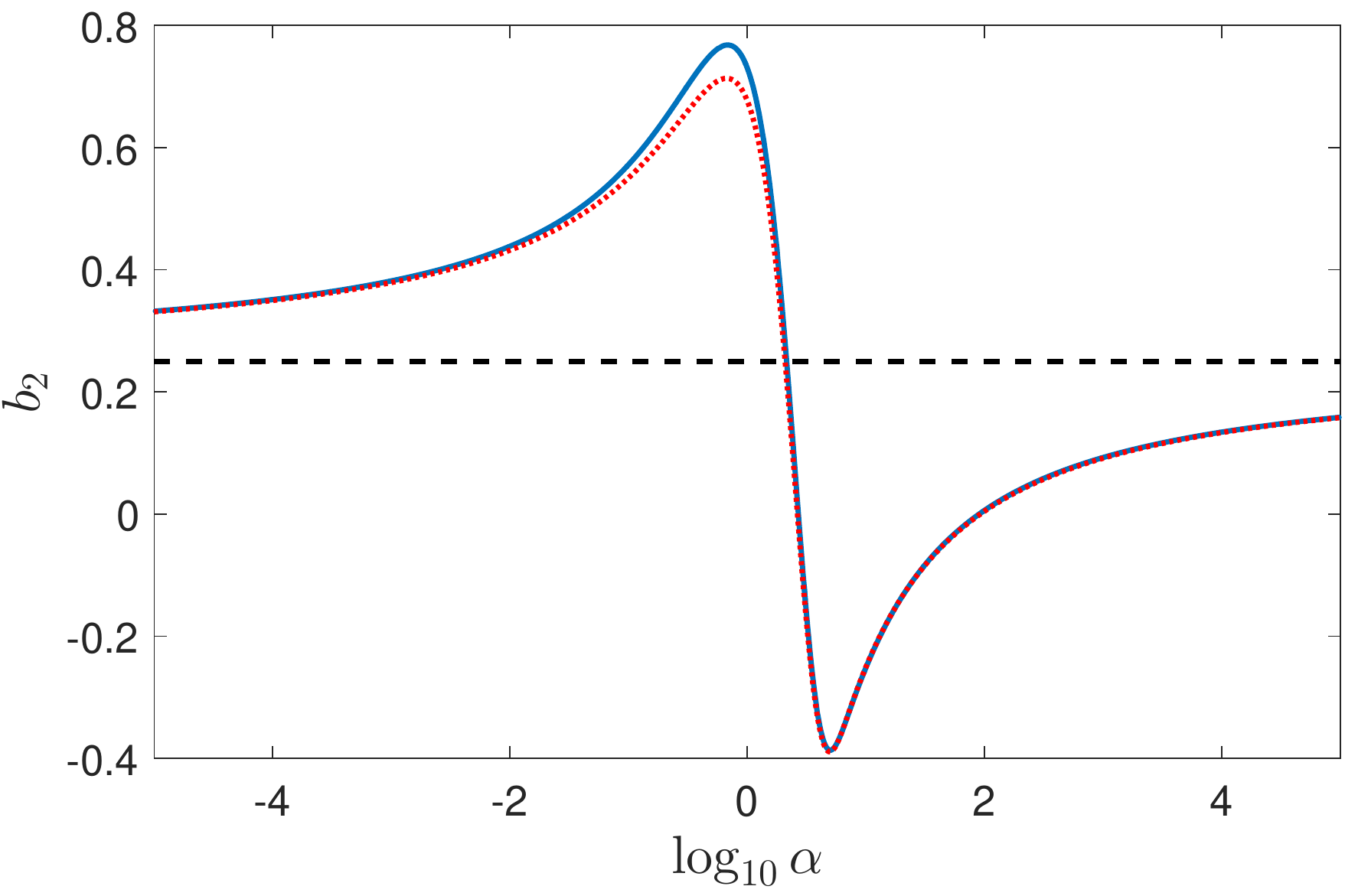}
\caption{The second virial coefficient $b_2$ as a function of $\alpha$ for bosons (solid blue).   
The dotted red curve is the approximation \eqref{b2stout}.    The non-interacting value 
$b_2 =1/4$  (dashed horizontal line)  is approached as $\alpha \to 0$ and $\alpha \to \infty$.}
\label{b2fig}
\end{figure}

The second virial coefficient $b_2$ is exact,  whereas $b_3$ is not since it does not contain the  intrinsic 
3-body physics. Hence we only consider $b_2$.        
Rescaling $\kvec \to \sqrt{2mT}\,  \kvec$,  and making the change of variables
$\kvec_1 = \kvec - \kvec'$,  $\kvec_2 =  \kvec + \kvec'$,  the integral factorizes and the 
integral over $\kvec_2$ is simply a gaussian.   The result is 
\beq
\label{b2}
b_2 =  \frac{s}{4}  +  \frac{2\sigma}{\pi}  \int_0^\infty  dk \, k\, e^{\frac{-k^2}{2}}  
{\rm arccot}  \[  \frac{2}{\pi}  \log \(  \frac{ \sqrt{\pi} \, k  }{\alpha} \) \]
\eeq
where $k = |\kvec|$ and $\sigma =1$ for bosons and $1/2$ for fermions.   
To a good approximation,  
\begin{align}
\nonumber
b_2 (\alpha) \approx  \frac{s}{4}  + \sigma \bigg( &  -1 +  2 e^{-\alpha^2/2\pi}+\ldots \\ \ldots -& \frac{2}{\pi}  
\arctan \[ \inv{\pi}  \log\( \frac{2 \pi \log 2}{\alpha^2} \bigg) \]  \).  
\label{b2stout}
\end{align}
Plots of $b_2$ are shown in Figure \ref{b2fig}.   Note that it changes sign 
at $\alpha\approx 2.65$.   As expected,  the free theory value $b_2 =s/4$ is approached
in both limits $\alpha \to 0$ and $\alpha \to \infty$.  

\clearpage

\end{document}